\documentclass[aps,prl,reprint,groupedaddress,amsmath,amssymb,longbibliography]{revtex4-2}

\usepackage{braket}
\usepackage{graphicx}
\usepackage{dcolumn}
\usepackage{bm}
\usepackage{xcolor}
\usepackage{xspace}

\begin{document}

\newcommand{\PECDlin}{$\mathrm {PECD}_\mathrm{lin}\ $}
\newcommand{\antiparallelstate}{$\ket {m_{\pm1}, H_{\mp}}$\xspace}
\newcommand{\parallelstate}{$\ket {m_{\pm1}, H_{\pm}}$\xspace}

\title{Sub-cycle resolved strong field ionization of chiral molecules and the origin of chiral photoelectron asymmetries}

\author{M. Hofmann$^1$}
\email{hofmann@atom.uni-frankfurt.de}
\author{D. Trabert$^1$}
\author{A. Geyer$^1$}
\author{N. Anders$^1$}
\author{J. Kruse$^1$}
\author{J. Rist$^1$}
\author{L. Ph. H. Schmidt$^1$}
\author{T. Jahnke$^2$}
\author{M. Kunitski$^1$}
\author{M. S. Schöffler$^1$}
\author{S. Eckart$^1$}
\author{R. Dörner$^1$}
\email{doerner@atom.uni-frankfurt.de}
\affiliation{$^1$ Institut f\"ur Kernphysik, Goethe-Universit\"at, Max-von-Laue-Str. 1, 60438 Frankfurt am Main, Germany\\
$^2$Max-Planck-Institut f\"ur Kernphysik, Saupfercheckweg 4, 69117 Heidelberg, Germany 
}
\date{\today}
\begin{abstract}
We report on strong field ionization of S- and R-propylene oxide in circularly polarized two-color laser fields. 
We find that the relative helicity of the two single color laser fields affects the photoelectron circular dichroism (PECD). 
Further, we observe that PECD is modulated as a function of the sub-cycle release time of the electron. 
Our experimental observations are successfully described by a heuristic model based on electrons in chiral initial states, which are selectively liberated by the laser field and, after tunneling, interact with an achiral Coulomb potential. 
\end{abstract}

\maketitle
Light-matter interaction is strongly determined by the symmetry properties of both the light field and the matter that is irradiated.
As a result, chiral molecules are particularly interesting because they lack many symmetries that are usually taken for granted \cite{Pitzer2013}, leading to many intriguing phenomena.
It has been shown that a circular dichroism can be observed in the photoelectron angular emission distribution upon ionization of chiral molecules. This effect was termed photoelectron circular dichroism (PECD) \cite{Ritchie1976,Bowering2001, Lux2012,Lux2015}. 
It manifests itself in a forward/backward emission asymmetry with respect to the light propagation direction.
Until now, the precise origin of PECD in strong field ionization remains debated: In one picture, signatures of molecular chirality are imprinted on the emitted electron by scattering in the long-range ionic potential after tunneling. 
Alternatively, those signatures could as well arise from the initial state and the subsequent tunneling dynamics, if the electron is e.g. liberated from a chiral molecular orbital. In this case, PECD momentum offsets already exist at the tunnel exit \cite{Beaulieu2016,Graefe2019,Bloch2021,Triptow2023}.
Experiments employing highly intense femtosecond laser fields to trigger tunnel ionization allow one to study PECD with attosecond time resolution \cite{Rozen2019,FehreElliptisch} and may provide an answer to that question.
In this Letter we use a co-rotating (CoRTC) and a counter-rotating (CRTC) circularly polarized two-color laser field to vary the effective angular frequency $\omega_\mathrm{eff}$ of the electric field vector during tunneling \cite{Eckart2018_Offsets}.
Three main findings will be presented: First, PECD depends on $\omega_\mathrm{eff}$. 
Second, the angle-resolved PECD differs when comparing electrons that are born on the rising and the falling edge of the laser field. 
Third, we find a radial momentum $p_\mathrm{r}$-dependent sign change of PECD. 
We use a heuristic model based on the strong field approximation (SFA) and a classical trajectory simulation (CTS) \cite{Geyer2022}, which shows that chiral initial bound states, in combination with a classical propagation of the electron after tunneling in an achiral Coulomb potential, are sufficient to describe the aforementioned results. In line with recent studies \cite{Triptow2023,Bloch2021}, this is consistent with chiral initial bound states being the main origin for the observed asymmetries \cite{Ordonez2022Propensity}.
\begin{figure}[th!]
\includegraphics[width=\columnwidth]{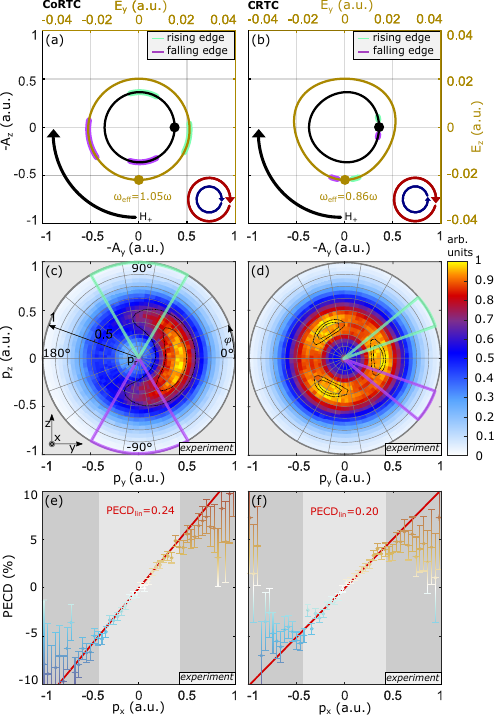}
\caption{\label{fig1} Strong field ionization of propylene oxide in two-color laser fields and the dependence of PECD on the effective angular frequency $\omega_\mathrm{eff}$. (a) [(b)] Lissajous curves of the CoRTC [CRTC] field $E(t)$ and the corresponding negative vector potential $-A(t)$. The helicity $H$ is indicated with a black arrow and is the same in (a) and (b). The red and blue arrows show the rotation of the laser electric field of the fundamental (red) and second harmonic (blue). The golden [black] dot marks $E(t_0)$  [$-A(t_0)$] at the time $t_0$ of maximal ionization probability. At this time $\omega_{\mathrm{eff}}$ is higher for CoRTC fields. (c) [(d)] Measured electron momentum distribution $p_\mathrm{final}$ in the polarization plane for the CoRTC [CRTC] field. The thin black lines are contour lines of regions with high intensity. The mint and purple circle segments highlight the angular gates on $p_\mathrm{final}$ which correspond to ionization times on the rising and falling edge of the laser field shown in (a) [(b)]. (e) [(f)] $p_{x}$-resolved PECD of the measured momentum distribution generated by the CoRTC [CRTC] field. The dark grey shaded areas are excluded from the calculation of $\mathrm {PECD}_\mathrm{lin}$. The PECD color scheme is the same as in Fig. \ref{fig2}.}
\end{figure}

The two-color laser fields are generated as the initial laser beam (40\,fs FWHM @ 790\,nm, 100\,kHz, Wyvern-500 by KMLabs) is divided and guided along two pathways, in one of which the second harmonic is generated ($200$-$\mu$m-$\beta$-barium borate crystal).
In each pathway of the interferometric two color setup $\lambda/2$- and $\lambda/4$-wave plates, as well as thin film polarizers control the polarization-state and intensity of the laser pulses. The relative phase between the two colors is set by a piezoelectric delay stage. Inside a cold target recoil ion momentum spectroscopy (COLTRIMS) reaction microscope \cite{Dorner2000,Jagutzki2002,Ullrich2003} we use a spherical mirror ($f=60$\,mm) to focus the pulses onto a molecular gas jet of S[R]-propylene oxide. After ionization, the charged fragments are guided by a homogeneous electric field of 47.3\,V/cm and a magnetic field of 7.8\,G to position- and time-sensitive detectors. 
Both, the electron and the ion arm had a length of 208.8 mm. 
The three-dimensional electron distribution was measured in coincidence with the singly charged parent ions to rule out fragmentation channel related effects  on PECD \cite{FehreAufbruch}. 

During the measurement the relative phase between the two colors was actively varied. These variations and additional slow drifts of the relative phase in the interferometer were compensated for in the offline data analysis as described in Ref. \cite{Eckart2018_Offsets}. 
The two-color laser pulses had an intensity of $5 \pm 2 \times 10^{13}\mathrm{W/cm^{2}} $ with a $\omega/2\omega$ field ratio of approx. 20/1. 
The intensity calibration was done \textit{in situ} for both colors separately:
The intensity of the fundamental and second harmonic light were deduced as in Ref. \cite{Hartung2019} and \cite{Eckart2016}, respectively.

Unlike circularly polarized light fields of a single color, CoRTC and CRTC fields vary within one laser cycle with regard to $|\vec{E}|$, $|\vec{A}|$ and $\omega$. 
Figure \ref{fig1}(a) [(b)] depicts the time-dependent CoRTC [CRTC] field  $\vec{E}$, the related negative vector potential $-\vec{A}$  and $\omega_{\mathrm{eff}}(t_0)=1.05 \omega$ [$\omega_{\mathrm{eff}}(t_0)=0.86 \omega$] for the used intensities in the interaction region, with being the instant of maximal ionization probability. 
The time dependence of two-color field leads to an angular dependence of the measured photoelectron momentum distribution in the polarization plane, as shown in Figs. \ref{fig1}(c) and \ref{fig1}(d). 
Building on the attoclock framework \citep{Eckle2008}, this mapping allows us to investigate PECD on subcycle time scales, which will be discussed in more detail later on. 

We define $\vec{y}$ and $\vec{z}$ to be perpendicular to the laser propagation direction $\vec{k_{\gamma}}\parallel\vec{x}$. 
The angle and radial momentum in the polarization plane are given as $\varphi=\mathrm{atan2}(p_z,p_y)$ and $p_r=\sqrt{p_y^2+p_z^2}$. 
For the sake of simplicity, we choose $\varphi=0^\circ$ to be at the highest intensity of the electron momentum distribution.
Thus, the coordinate frames in Figs. \ref{fig1}(a) and \ref{fig1}(b) are rotated with respect to the ones in Figs. \ref{fig1}(c) and \ref{fig1}(d) by the influence of the Coulomb potential \cite{Bray2018}. 

In case of PECD induced by circularly polarized light of a single color, swapping the light's helicity is fully equivalent to swapping the molecular handedness because of the cylindrical symmetric photoelectron distributions in the polarization plane. However, this is not the case for two color fields. In order to properly account for the angular dependency of the two-color field, we hence define: 
\begin{align}\label{PECD}
\mathrm {PECD} = \frac{A-B} 
{A+B} 
\end{align}
\noindent with
\begin{align}\label{PECDABCD}
A  = I_{R,H_{+}}+I^*_{S,H_{-}},\  
B  = I_{S,H_{+}}+ I^*_{R,H_{-}}
\end{align}
\noindent where * denotes the data sets which are mirrored at the $p_{x} p_{y}$-plane. 
$I$ is the intensity for a given enantiomer $R$ or $S$ and overall light helicity $H_{+}$ or $H_{-}$. 
$I$ represents the three dimensional momentum distribution and thus is a function of $p_{x}$, $p_{y}$ and $p_{z}$.

As PECD depends approximately linearly on $p_{x}$,
\begin{align}\label{PECDlin}
\mathrm {PECD}_{\mathrm{lin}}  = \frac{\partial}{\partial p_{x}} \mathrm {PECD}
\end{align}
is used to quantify PECD by a single number \citep{FehreElliptisch}. It is evaluated for those values of $p_{x}$, where PECD dependency is approximately linear (light grey areas in Figs. \ref{fig1}(e) and \ref{fig1}(f)).
\begin{figure}[th!]
\includegraphics[width=\columnwidth]{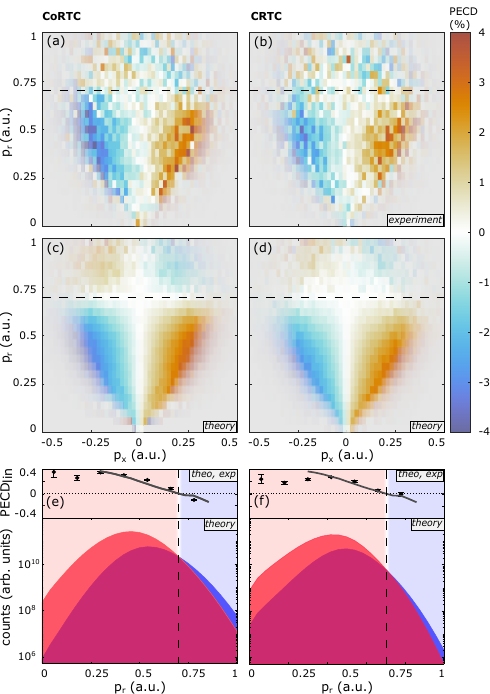}
\caption{\label{fig2} 
Momentum resolved PECD. 
Measured (a) [(b)] and simulated (c) [(d)] PECD as a function of $p_x$ and $p_r$ for the CoRTC [CRTC] field. 
The opacity of each bin indicates the intensity of the measured 
photoelectron momentum distribution. 
PECD decreases for rising $p_r$-values and changes its sign at $p_{r}\approx 0.7$ a.u.
(e) [(f)] Final momentum distributions obtained from \antiparallelstate (red) and \parallelstate (blue) for the CoRTC [CRTC] field.
Additionally, the PECD sign change along $p_r$ coincides with the relative change in the contributions of \antiparallelstate and \parallelstate. The simulated (grey lines) and measured $p_r$-resolved \PECDlin spectra are shown on top.}
\end{figure}
In order to investigate the dependence of the chiral response on the effective frequency $\omega_\mathrm{eff}$, \PECDlin is calculated from the data presented in Figs. \ref{fig1}(c) and \ref{fig1}(d).
The $p_{x}$-resolved PECD of propylene oxide shown in Fig. \ref{fig1}(e) [(f)] for CoRTC [CRTC] as well as the \PECDlin exhibit a significant dependence on the angular velocity of the electric field vector: 
An increase of $\omega_\mathrm{eff}$ by 20\% leads to an increase of \PECDlin of about 20\%. 
For CRTC light the second harmonic counteracts the angular frequency of the fundamental, reducing the chiral response. 
This is in line with the approximately linear dependence of PECD on the ellipticity of the light for elliptically polarized light \cite{FehreElliptisch}. 

We now focus on the measured $p_r$-resolved PECD shown in Fig. \ref{fig2}(a) and  \ref{fig2}(b). 
Overall, for both the CoRTC and CRTC field, the $p_x$-dependency and the \PECDlin follows the same trend as discussed in the $p_r$-unresolved case above (Figs. \ref{fig1}(e) and \ref{fig1}(f)). 
Nevertheless, this dependency changes along $p_r$: The PECD $p_x$-asymmetry decreases ($0.3\ \mathrm{a.u.} <p_r<0.7\ \mathrm{a.u.}$) to a point, where it  flips ($p_r>0.7$ a.u.).
This sign change is in line with experimental \cite{FehreAufbruch} and theoretical \cite{Artemyev2022} strong field PECD studies and can be interpreted as a result of chiral propensity rules caused by chiral initial bound states of liberated electrons \cite{Ordonez2022Propensity}. 

To gain insight into the underlying physical mechanisms, we employ a heuristic SFA+CTS model, based on the idea of those chiral initial states.
Our model uses the following assumptions:
(i) There are initial bound states that possess a nonzero angular momentum and hence an angular dependent phase in coordinate space in a certain plane (that can be chosen for a given molecule).
In the simplest case this would be an atomic p-state with a magnetic quantum number of $m=\pm1$ with $\vec{k_{\gamma}}\parallel\vec{x}$ being the quantization axis. For a given light helicity $H$, the sense of rotation of the phase of the p-state in position space is either parallel (\parallelstate) or antiparallel (\antiparallelstate) to the sense of rotation of the laser electric field.
(ii) The m-selectivity of the tunneling \cite{Smirnova2011, Smirnova2013, Smirnova2015, Li2015, EckartRingstrom}, which is accounted for in SFA \cite{OlgaSFATheorie}, ensures that the strong field ionization rate of the \antiparallelstate-state 
 is enhanced, and, consequentially, depleted for the \parallelstate-state.
(iii) In a chiral molecule, these initial states cause a slight momentum offset in or against the direction of light propagation after tunneling depending of the enantiomer. For one enantiomer, the direction of this offset is firmly associated with the sign of m (e.g. ``backward displacement" for $m=-1$ (see Fig. \ref{fig1} in \cite{Ordonez2022Propensity})) and flips for the other enantiomer.
(iv) In the continuum, the electron propagates classically. Its dynamics are solely governed by the laser field and an achiral Coulomb potential. 
\begin{figure}[!b]
\includegraphics[width=\columnwidth]{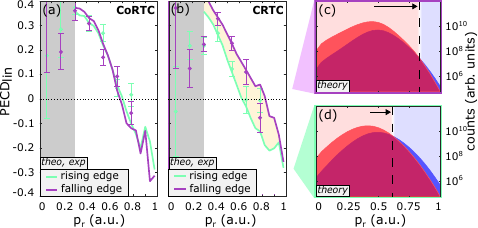}
\caption{\label{fig3} 
Subcycle-resolved PECD. 
(a) [(b)] Measured and simulated $p_r$-resolved \PECDlin of the rising and falling edge of the CoRTC [CRTC] field. 
The falling edge of CRTC field shows an increased PECD compared to the rising edge, which is reproduced by our SFA+CTS simulation. 
As in Fig. \ref{fig2} this difference can be linked to the corresponding contributions of \parallelstate (blue) and \antiparallelstate (red) to the final momentum distribution, which is illustrated in (c) [(d)] for the falling [rising] edge of the laser cycle.
For small $p_r$ values (grey shaded areas) the simulation deviates from the measurement due to an overestimation of Coulomb focusing and is therefore not shown in (a) and (b).}
\end{figure}

Within our model, the tunneling rates are calculated using the SFA for a hydrogen-like atomic state. The magnitude of the photoelectron momentum offset after tunneling along the light-propagation direction is chosen to be $\Delta p_x =\pm 0.006$ a.u. to obtain the best agreement of experiment and simulation. We emphasize that $\Delta p_x$ is the only external parameter. 
\begin{figure*}[]
\includegraphics[width=\textwidth]{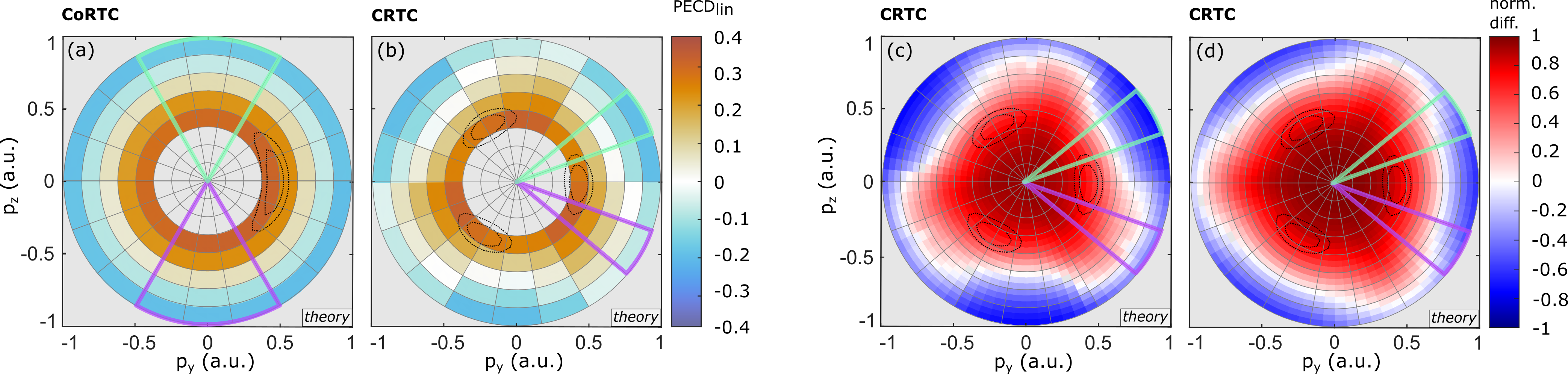}
\caption{\label{fig4} 
Influence of the long range Coulomb potential on the PECD asymmetries. 
(a) [(b)] Simulated sub-cycle resolved \PECDlin of CoRTC [CRTC] light fields in the polarization plane. 
(b) exhibits a difference in \PECDlin comparing the rising and falling edge of the laser field seen in Fig. \ref{fig3}(b). 
(c) Normalized difference of the final momenta of \parallelstate and \antiparallelstate in the polarization plane. In accordance to Figs. \ref{fig2}(e), \ref{fig2}(f) and \ref{fig3}(c), \ref{fig3}(d), \antiparallelstate [$\ket {m_{\pm1}, H_{\pm}}$] is dominant in the red [blue] area.
Comparison with (b) underlines the connection between those initial states and PECD. 
(d) The same as (c) but without the influence of the long range Coulomb potential. 
The latter leads to a nonlinear mapping of the chiral initial momenta after tunneling to the final momentum space. 
Neglecting the long range Coulomb potential restores the symmetry of the normalized difference when comparing the rising and falling edge of the laser field.
}
\end{figure*}
For the purpose of distilling the physics underlying the observed photoelectron emission asymmetry, the undercomplex representation of the states of a chiral molecule in terms of hydrogen-like achiral ones and m-selective chiral momentum offsets after tunneling is a clear strength. One price to be paid for this simplistic but transparent modelling is that it does not have predictive power on the sign of that asymmetry. It also doesn't answer the question whether the offset is a caused by the initial state wavefunction or is imprinted onto the photoelectron flux during tunneling. For the sake of simplicity, all calculations will be shown for a clockwise rotating light field $H_\mathrm{+}$ as in Fig. \ref{fig1}.

Figs. \ref{fig2}(e) and \ref{fig2}(f) each show the simulated $p_r$-resolved final momentum distributions for \parallelstate (blue) and \antiparallelstate (red). 
In line with Ref. \cite{Ordonez2022Propensity}, \antiparallelstate [$\ket {m_{\pm1}, H_{\pm}}$] is liberated more [less] efficiently from the bound state and its final momentum distribution is centered around lower [higher] radial momenta.
As a result, electrons with a radial momentum $p_{r}< 0.7$ a.u. stem predominantly from \antiparallelstate, while for higher $p_{r}$, the electrons originate mostly from \parallelstate. 
Consequentially, due to point (iii) of our model, this final momentum composition along $p_{r}$ exactly reproduces the $p_r$-dependency of the PECD and the characteristic sign change of the PECD comparing electrons below and above $p_{r}\approx 0.7$ a.u. Accordingly, the simulated PECD distributions depicted in Figs. \ref{fig2}(c) and \ref{fig2}(d) are in good agreement with our measured results, see Figs. \ref{fig2}(a) and \ref{fig2}(b).
Furthermore, our model can also explain, why PECD is stronger for the CoRTC field as compared to the CRTC field. This is because the higher effective angular frequency $\omega_\mathrm{eff}$ of the CoRTC field leads to a more selective liberation of \parallelstate and \antiparallelstate compared to the CRTC field-case, which has a lower effective angular frequency $\omega_\mathrm{eff}$ (see Fig. \ref{fig2}).


In order to study PECD in even greater detail, Fig. \ref{fig3} 
resolves the PECD's sub-cycle dependency. 
We compare the PECD for cases where we set angular gates (as indicated in Fig. \ref{fig1}) which correspond to the rising and falling edge of the electric field within a laser cycle. 
Here, a significant difference between PECD in CoRTC [Fig. \ref{fig3}(a)] and CRTC [Fig. \ref{fig3}(b)] fields can be found. 
This difference is reliably predicted by our heuristic model. 
Examination of the contribution of \parallelstate and \antiparallelstate to the electron momentum distribution shown in Fig. \ref{fig3} (c) and (d) indicates the physical origin of the observed difference in sub-cycle resolved PECD seen in Figs. \ref{fig3}(a) and \ref{fig3}(b): The rising-edge-gated electron distribution is more strongly built up by \parallelstate initial states and thus shows a decreased (less pure) PECD signal. 
Additionally, the PECD's sign change is consistently shifted to smaller values of $p_r$.
The deviations comparing the rising and the falling edge that are illustrated in Fig. \ref{fig3} are particularly interesting, since the properties of the laser field ($E(t)$, $A(t)$ and $\omega_\mathrm{eff}$) at the instant of tunneling 
are very similar, as the gates are symmetrically chosen around the field maximum of the laser cycle [see Fig. \ref{fig1}(a) and \ref{fig1}(b)]. 

To further elucidate on that, we finally use our model to understand the influence of the long-range Coulomb potential on the observed PECD asymmetries: 
Fig. \ref{fig4}(a) [(b)] shows the fully resolved \PECDlin distributions as Fig. \ref{fig3}(a) [(b)] 
in the polarization plane. The differences comparing the rising and the falling edge of the CRTC field are reproduced [see angular gates in Fig. \ref{fig4}(a) and \ref{fig4}(b)].
Next, we visualize the respective contributions of the chiral initial states to the final electron momentum distribution in the polarization plane. For this the normalized difference of \antiparallelstate and \parallelstate is plotted in Fig. \ref{fig4}(c).
It is evident that this composition mimics the distribution of \PECDlin as expected, highlighting the deep connection of PECD and the chiral initial states. 
Finally, we artificially switch off the Coulomb potential after tunneling in our simulation. In full analogy to Fig. \ref{fig4}(c) this leads to Fig. \ref{fig4}(d). Strikingly, neglecting the Coulomb potential restores the symmetry of the normalized difference with respect to the rising and falling edge of the laser field.
We therefore conclude that the Coulomb interaction leads to an $\varphi$-dependent distortion of the final momentum distributions, but does not create a chiral sub-cycle asymmetry by itself. 
This nonlinear initial- to final-momentum mapping ultimately results in the observed differences of the PECD when comparing electrons born on the rising and the falling edge of the laser field.

In conclusion, we have measured the three-dimensional electron momentum distribution for strong field ionization of S- and R-propylene oxide using CoRTC and CRTC laser fields. 
We find that, first, the overall PECD is stronger in CoRTC fields than in CRTC fields. 
This finding highlights the importance of the effective angular frequency $\omega_\mathrm{eff}$ of the laser electric field for chiral light-matter interaction in the strong field regime \cite{Eckart2018_Offsets, Ordonez2022Propensity} as it affects the inital-state selectivity of the ionization process. 
Second, in both cases, the radial momentum $p_\mathrm{r}$-dependent sign change of the PECD is apparent. 
Third, we have analyzed PECD with sub-cycle resolution. We find that PECD is different for electrons that are released on the rising and the falling edge of the CRTC laser cycle. 
The success of a heuristic SFA+CTS model using an achiral potential and a simplified chiral initial electronic bound state is consistent with chiral initial bound states being the main  reason for the measured chiral photoelectron asymmetries, while the long-range Coulomb potential only induces an achiral $\varphi$-dependent distortion after tunneling \cite{Eckle2008,Trabert2021atomicH,Ordonez2022Propensity}. 
Our data shows that this explanation even holds for fully three-dimensionally resolved photoelectron momentum distributions. Thus, our work motivates future PECD studies with even more complex laser fields.

\section{Acknowledgments}
\normalsize
The experimental work was supported by the DFG (German Research Foundation). M.H. acknowledges funding of the DFG through CRC 1913 ELCH. We thank Manfred Lein and Simon Brennecke for insightful discussions on SFA simulations.

%

\end{document}